*Molecular mechanism of water permeation in helium impermeable graphene and graphene oxide membrane*

*Nallani Raghav, Sudip Chakraborty and Prabal K Maiti*[*]

Center for Condensed Matter Theory, Department of Physics, Indian Institute of Science,

Bangalore 560012

ABSTRACT: The layers of graphene oxide (GO) are found to be good for permeation of water but not for helium (*Science* **2012** 335 (6067): 442-444) suggesting that the GO layers are dynamic in the formation of permeation route depending on the environment they are in (i.e, water or helium). To probe the microscopic origin of this observation we calculate the potential of mean force (PMF) of GO sheets (oxidized and reduced parts), with inter-planar distance as reaction coordinate in helium and water. Our PMF calculation shows that equilibrium interlayer distance between oxidized part of GO sheets in helium is at 4.8 Å leaving no space for helium permeation. In contrast PMF of oxidized part of GO in water shows two minima one at 4.8 Å and another at 6.8 Å corresponding to no water and water filled region and thus giving rise to permeation path. The increased electrostatic interaction between water with the oxidized part of the sheet helps the sheet opening up and pushing water inside. Based on the entropy calculations for water trapped between graphene sheets and oxidized graphene sheets at different inter-sheet spacing we also show the thermodynamics of filling.

[*] maiti@physics.iisc.ernet.in

**Introduction**

Single sheet of graphene or graphene oxide (GO) is one atom thick and impermeable to most other molecules [1]. These sheets, particularly GO, have been produced in the form of stacked layers [2]. Such layered structures are of considerable interest as they show remarkable ability to differentiate between molecules [3]. The immediate application that comes to mind is that of filtration of unwanted ions from water [1, 4, 5]. Also properties of water confined inside narrow channel of carbon nanotube or in the slit pore geometry of two graphene or graphene oxide sheets (GO) have received lots of interest in recent years [6-10]. Confined water in these environment exhibits properties distinct from the bulk water. Demonstrating the ability of GO to differentiate molecules, Geim et. al. [11] showed that water was able to find permeation routes through these stacked layers of GO as opposed to helium, which showed negligible permeation rate. Helium, given its size and negligible interaction with substrate is an important gas that is regularly used in leak testing of membranes and thin films [1]. If a membrane is helium leak tight, it is assumed to be impermeable to most other gases and liquids. So the unimpeded permeation of water through helium leak tight membranes challenges the conventional wisdom about helium leak properties and demands an immediate explanation. In this letter we propose a mechanism which explains how GO membrane allows water permeation although being a helium leak proof. To support our proposed mechanism we first develop atomic model of GO and use all atom MD simulations for calculating free energy of GO using force integration method and 2 phase thermodynamic method to compute free energy of water trapped inside the membrane. Our proposed mechanism may also open up the possibility of designing novel helium leak proof membrane and better testing mechanism of membrane leakage.

Recently, there is also a huge demand to design devices that store helium without moisture. Inhaled helium-3 helps in giving good MRI picture of the lungs [12], but to reuse the helium gas that is exhaled, it should be devoid of moisture. By being helium impermeable in nature, GO can help in making such a recycling device as it allows water to permeate across itself freely. The extreme sensitivity of water to the GO is also being investigated for designing humidity sensors [13]. We believe that our work will have considerable impact in design and innovation of devices involving GO as it provides an insight into dynamic pore formation. Recently Xu et. al. [14] have studied the selective gas permeation through the space between GO sheets using MD simulation. In another related study [10] they also showed that hydrodynamics needs to be included to account for the water dynamics in the channel between GO sheets.

To support our proposed mechanism we present our study in two parts. In the first part we study the permeation properties of reduced parts of GO by using pristine graphene. In the second part we study the permeation behaviour of the oxidized parts of GO using functionalized graphene. We present Potential of Mean Force (PMF) calculations of pristine graphene and functionalized graphene (at 13.5% and 33% of oxidation) in water[15, 16] and helium using force integration. The level of oxidation for the functionalized graphene used in our work was motivated by the recent reaxFF simulation of GO as well experimental observation [17, 18]. We also present entropy and free energy of water trapped between the reduced part of the GO membranes (using pristine graphene) and oxidized parts of GO membranes using functionalized (33%) graphene sheets using 2 phase thermodynamics (2PT) method [19, 20] to elucidate our proposed mechanism. The 2PT method has been shown to be very reliable in estimating entropy of water under variety of conditions: entropy of water inside CNT [7, 21, 22], entropy of water in the hydration layer of bilayer [23, 24], DNA and dendrimer [25, 26], entropy of various organic liquids [27].

**Simulation Details**

We have used 2 parallel pristine graphene sheets each containing 84 carbons and solvated them in 1780 molecules of TIP3P water [28]. Amber FF03 [29] was used to model the interaction of carbon of pristine graphene (atom type CA) which has been earlier shown to describe structure and dynamics of confined water very well [30, 31]. The system was minimized by both 500 steps of steepest descent and 500 steps conjugate gradient methods to remove the bad contacts between the sheet and water. During minimization the pristine graphene sheets were held fixed to their initial positions using a harmonic constrain with a spring constant of 500 kcal/mol-Å$^2$. Then the system was slowly heated restraining the solute using a harmonic constrain with a spring constant of 10 kcal/mol-Å$^2$ from 0K to 300K in steps of 30 K for each 10 ps using Langevin thermostat. The sheets were held parallel to each other by applying multiple force constants at various points on the sheets and were equilibrated at 1 atm pressure and at 300 K using Nose Hoover thermostat and barostat at different inter-sheet spacing for 1 ns. We used a time constant of 0.1 ps and 2 ps for thermostat and barostat respectively. Velocity Verlet scheme was used for integrating the equation of motion with a time step of 2 fs. Particle-Particle Particle-Mesh Ewald summation method (PPPM) [32] was used to compute the long range part of the electrostatic interaction with a tolerance of 0.0001. 10Å cut-off was used to compute the short-range LJ interaction as well as the short range part of the Coulomb interaction. Same sheets were solvated in helium in the same way. We have used the helium interaction parameter from CHARMM [33] to describe the interaction between the carbon and helium. Same helium parameters were used recently [34] to study the stability of helium bubble in water. To do the simulation at liquid state of helium we have done the helium simulations at high pressure (2700 atm). The system had 2000 helium atoms and was equilibrated at 300K.

We have developed the molecular model of functionalized graphene as follows: epoxy bonds and hydroxy bond are grafted to the graphene sheet on both sides randomly [17, 18, 35]. The functionalized graphene sheet was optimized through Gaussian and ESP charges were calculated. The Gaussian optimized structure with the ESP charges was used in ANTECHAMBER module of AMBER [36] to derive the GAFF parameter [37] and RESP charges for the subsequent MD simulation. The RESP charges and atom types are shown in Figures S1 and S2 in the supplementary materials. The functionalized graphene sheets were minimized and equilibrated similarly as the graphene sheets in 6000 TIP3P water molecules for water case and 15000 helium atoms for helium case as shown in Figure S3. The helium case was done at the same pressure as the pristine graphene in helium system (at 2700 atm) corresponding to the liquid state.

Potential of Mean Force calculations were done using force integration, with a window size of 0.25Å as follows

$$PMF(d) = \int_{\infty}^{d} f(r)\,dr$$

Where $d$ and $r$ are the distances between the sheet's center of mass, perpendicular to the plane of the sheets, $f(r)$ is the force on a sheet due to the other sheet and solvent.

The level of oxidation for the functionalized graphene used in our work was motivated by the recent reaxFF simulation of GO as well experimental observation. At each window system was equilibrated for 5 ns continuing from the previous windows and data was collected for the last 2 ns of the production run. The system consists of two sheets (both pristine graphene or both functionalized graphene) and the solvent (water or helium). The sheets were kept parallel to each other using multiple force constants.

The entropy of water molecules are calculated using 2PT method. The details of the 2PT methods can be found in the original papers [19, 20, 38-40]. Here we briefly mention the major steps in the 2PT method. In this method the Fourier transform of the velocity auto-correlation function, obtained from an MD run is used to obtain density of states (DoS), which is then used in the calculation of the thermodynamic properties by applying quantum statistics to each mode, assuming that it can be described as a harmonic oscillator. This process works well for a solid but it is not correct for a liquid as liquids have finite DoS at zero frequency, leading to a finite value of the diffusion constant. The problem is resolved by using a two-phase model for the liquid, comprising of a solid phase, the DoS for which goes to zero smoothly and a fluidic gaseous phase, described as a gas of hard spheres. The dynamical information of the water molecules (both in bulk and confined), can be extracted from the vibrational spectrum of the water molecules, determined from the Fourier transform of the velocity auto-correlation function $S(\upsilon) = \frac{2}{k_B T} \lim_{\tau \to \infty} \int_{-\tau}^{\tau} C(t) e^{-i2\pi \upsilon t} dt$ where C(t) can either be mass weighted velocity auto-correlation function determined from the centre of mass velocities $v^{CM}(t)$ of the water molecules $C_T(t) = \sum_{i=1}^{n} \langle m_i v_i^{CM}(t) \cdot v_i^{CM}(0) \rangle$ or the moment of inertia weighted angular velocity auto-correlation function

$C_R(t) = \sum_{j=1}^{3} \sum_{i=1}^{n} \langle I_{ij} \omega_{ij}^{CM}(t) \cdot \omega_{ij}^{CM}(0) \rangle$ where $I_{ij}$ and $\omega_{ij}(t)$ are the moment of inertia and angular velocity tensor of the water molecule *i*. For the computation of entropy, the DoS is partitioned into a solid like and gas like component using a fluidicity factor *f* which is a measure of fluidity of the system.

## II. RESULTS AND DISCUSSION

The PMF profiles of graphene in water and in helium are shown in Figure 1. Both profiles show a global minimum at 3.4 Å which is the closest the sheets can be. This implies that the reduced parts of GO tend to close up whether they are in water or helium environment. In water, the effective interaction between the pristine graphene is very large due to the strong vdW interaction giving a barrier height of 180 kcal/mol. Such large barrier height for PMF profile of graphene sheets in water is very similar to other available literature values [15]. Another feature of the PMF of graphene in water is that it has local minima at 6.75 Å and 9.5Å distance between the sheets. These correspond to single and double layers of water respectively. These two minima are separated by low energy barrier of 20 kcal/mol. The instantaneous snapshots of the graphene sheets with single and double layers of water are also shown in Figure 1.

The PMF profiles of functionalized graphene for 13.5% and 33% oxidation in water shown in Figures 2 and 3 respectively, have two global minima one at around 4.8Å another at around 6.8Å separation of the sheets. The first minimum corresponds to situation where there is no water between the functionalized graphene sheets. The second minimum at around 6.8Å corresponds to one layer of water between the sheets. It is worth mentioning here that recent scanning force microscopy (SFM) on the hydrated GO bilayer[41] reported similar interlayer distance increase when immersed in liquid water. In both the oxidation cases (13.5% as well as 33%) these minima (corresponding to no water layer and single layer of water) are separated by free energy barrier of the order of 5 kcal/mol or less [16]. This is in contrast to the pristine graphene case (figure 1) where the free energy barrier between the no water and single water configuration is typically [15] of the order of 100 kcal/mol. So the functionalized graphene layers are more prone to have single layer of water between the sheets. Another feature to note from this profile is the reduced free energy barrier between global minima at 4.8Å and almost flat profile beyond 9 Å. The flat profile beyond 9 Å corresponds to almost bulk like water between the sheets and the reduced barrier which is of the order of 3 kcal/mol can be easily breached. This suggests that the water can easily get accumulated between the oxidized parts of a GO membrane thereby wedging open the membrane. It is worth mentioning here that Lerf et. al. [42] using neutron scattering study also reported a layer spacing of 8, 9 and 11 Å depending on the relative humidity level. Level of humidity also controls the permeation rate. Nair et. al. [11] showed that increasing the humidity level allow the permeation of He through the GO membrane which is otherwise impermeable. However, in a recent work Jiao and Xu [43] shows that increasing humidity level lower the permeation of various gasses.

In contrast the PMF profiles of functionalized graphene in helium shown in Figures 2 and 3 are very much similar to the case of pristine graphene in helium with only one global minimum at the minimum inter-sheet spacing of 4.8Å. This finally gives us the picture of non permeability of helium through GO. Both reduced and oxidized parts of layered GO having their free energy minimum at closest distance of 3.4Å and 4.8Å respectively to another sheet, refuse to provide permeation pathway as it happens at a steep price of large free energy barrier of 120 kcal/mol and 70 kcal/mol respectively.

To have further microscopic origin of the above mentioned behaviour we also calculate the entropy and free energy of a water molecule confined between reduced parts of GO sheet (pristine graphene) as well as oxidized parts of a GO sheet using 2PT method. From the 2PT calculations, we see that the entropy of single layer of TIP3P water between reduced parts of GO (at a separation of 6.5Å, Table I.) are higher compared to that of single layer of water between oxidized parts of GO (at a separation of 7.0Å, Table II.) This is due to the better electrostatic interaction between the oxidized parts of sheets and water. More negative interaction energy for water (-10 kcal/mol) trapped inside GO sheet at 7.0Å separation compared to the pristine graphene (-9 kcal/mol) as shown in Table I and Table II, demonstrate this. Even for more layers of water the comparison holds, because the surface of membrane, if it is an oxidized one, locks the top layer of water with its electrostatic interaction, thereby reducing its mobility which results in locking of the subsequent layers of water too.

This is also validated by the diffusion coefficient of water between oxidized and reduced parts of GO as shown in table 1 and table 2. These results corroborate qualitatively with the recently reported flow enhancement values calculated using non-equilibrium molecular dynamics simulations by Xu et. al. [14]. Also from the average number of hydrogen bonds per water molecule (Table S1 and S2 of the SI), we see that at every separation, water trapped between pristine sheets have higher number of hydrogen bonds than for water trapped between oxidized parts. This is due to superior sheet water interaction (in the oxidized case) that pulls water layers slightly apart and thus reducing the number of Hbonds per water. This is again similar to the finding of Xu et. al. [14] where they show that the edge-pinning effect breaks down the ultrafast flow of water within the pristine graphene gallery sandwiched between oxidized regions in graphene oxide sheets. The hydrogen bonds between water molecules in the pristine and oxidized regions also play a central role in controlling the sheet separation and their mechanical properties [17, 18].

The free energy on the other hand except for single layer between the reduced parts of GO sheets, is almost similar across all the separations of oxidized parts of GO and reduced parts of GO. Lower internal energy for water between the reduced parts of GO is compensated by high entropy, to the high internal energy gained by water trapped between the oxidized parts of GO from interaction with the sheet. From this we can conclude that once there is an opening wide enough for more than a single layer of water between the oxidized parts of GO or reduced parts of GO sheets, the water is free to move in and out of the reduced and oxidized areas. This is similar to the case of water entry into the nanotube, where the free energy is similar to bulk and inside [7]. Here the slit pore is dynamic in nature (as opposed to constant pore size of a nanotube), and is formed only when the oxidized parts open up (as suggested by our PMF calculations of functionalized graphene). This suggests that the oxidized parts are used to open the layer spacing, but the diffusion happens only between the reduced parts of the layers.

## III. CONCLUSION

The following microscopic picture emerges from our calculation. In the case of GO, water stuck in between reduced parts of GO (pristine graphene) will be squeezed out because the loss in free energy for pristine graphene sheets by letting a single layer of water coming between them is not compensated by gain in entropy of water coming from bulk environment. In the case of oxidized parts of GO, the free energy change will be negligible for the sheets in going from their minimum distance separation (corresponding to no water in between) between them to a single or more layers of water configuration trapped

in- side. So for a molecule of water in the bulk it would look for the oxidized parts of the layer to aggregate into. Once there is bulk like environment between the oxidized parts of a layer, the membranes are wedged open resulting in opening of reduced parts too as shown in the schematic Figure 4. This is in contrast to an oxidized nanotube where after accumulation of water between the oxidized parts there cannot be any wedge opening leading to reduced permeation. The water that enters these reduced parts (pristine graphene part) of a layer due to negligible change in the free energy will promptly be squeezed out to bulk or oxidized parts, thereby creating channels of permeation. We have also tested our proposed mechanism through an actual simulation. Snapshots of a representative configurations of this MD simulation of a mixed functionalized graphene sheets (60% of functionalized graphene for 33% oxidation and 40% of its reduced counterpart with a total area of 5 nm x 5 nm) is shown in figure 4(b). As proposed in our schematic, we see the opening of a capillary in the oxidized parts of the membrane in the presence of water.

ACKNOWLEDGMENTS

We thank DST, India for funding and Prof. Manish Jain for helpful suggestions.

FIGURES

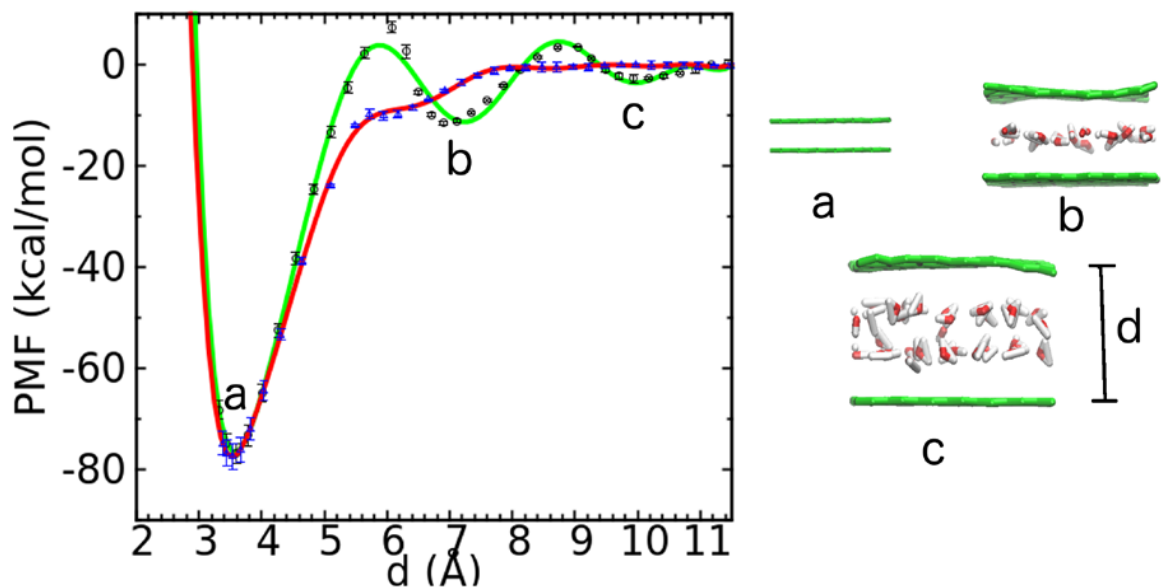

Figure 1: PMF of pristine graphene in water (green) and in helium (red). The configurations of graphene in water corresponding to global minima at (a) 3.4Å and (b) at 7Å and local minimum (c) at 9Å are shown on the side. Polynomial fit to the PMF profile is a guide to the eye only

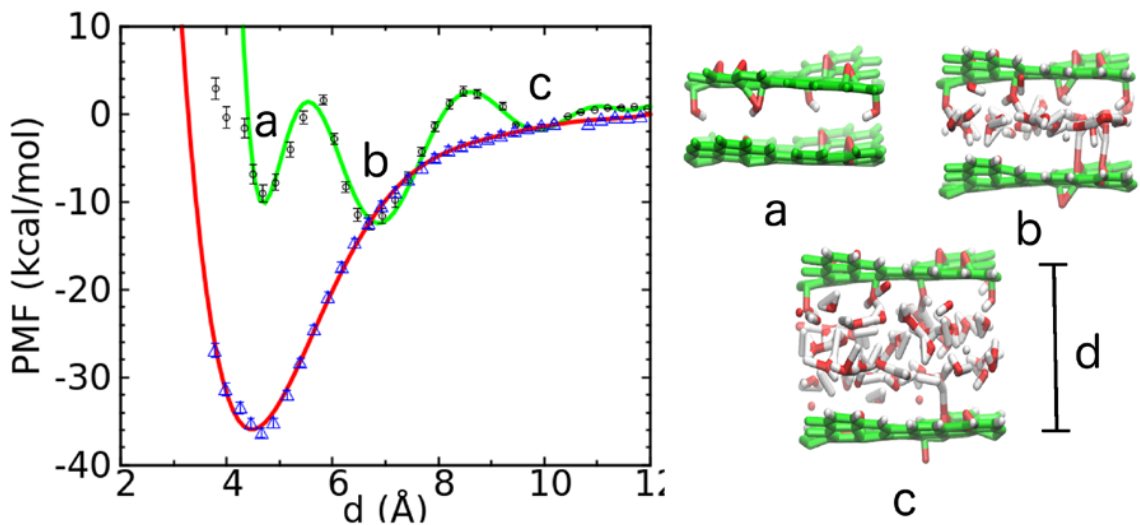

Figure 2. PMF of functionalized graphene (13.5% oxidation) in water (green) and helium (red). The configurations of functionalized graphene in water corresponding to global minima at (a) 4.8Å and (b) at 7Å and local minimum at (c) 10Å in water are shown on the side. Polynomial fit to the PMF profile is a guide to the eye only

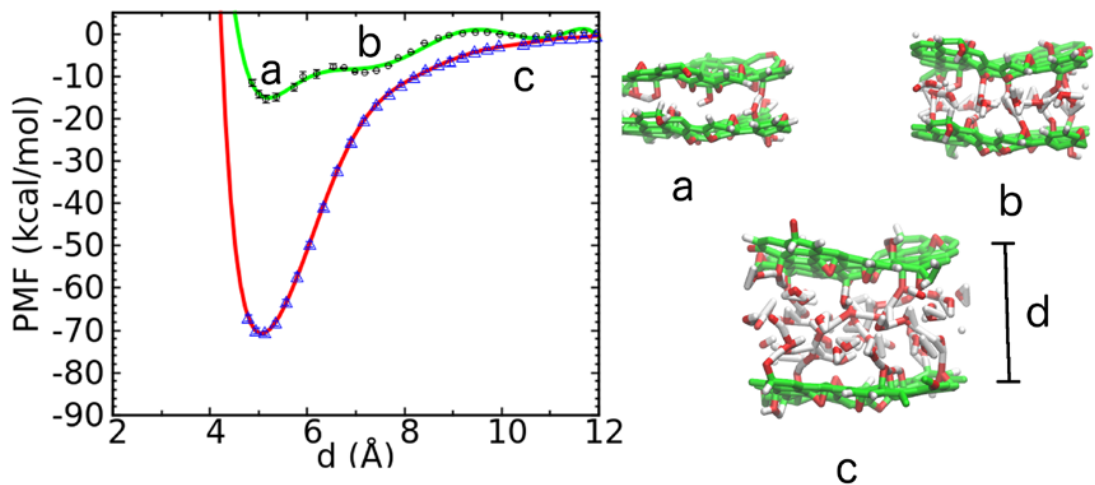

Figure 3: PMF of functionalized graphene (33% oxidation) in water (green) and helium (red). The configurations corresponding to global minima at (a) 4.8Å and (b) at 7.0Å and local minimum at (c) 10Å are shown on the side Polynomial fit to the PMF profile is a guide to the eye only.

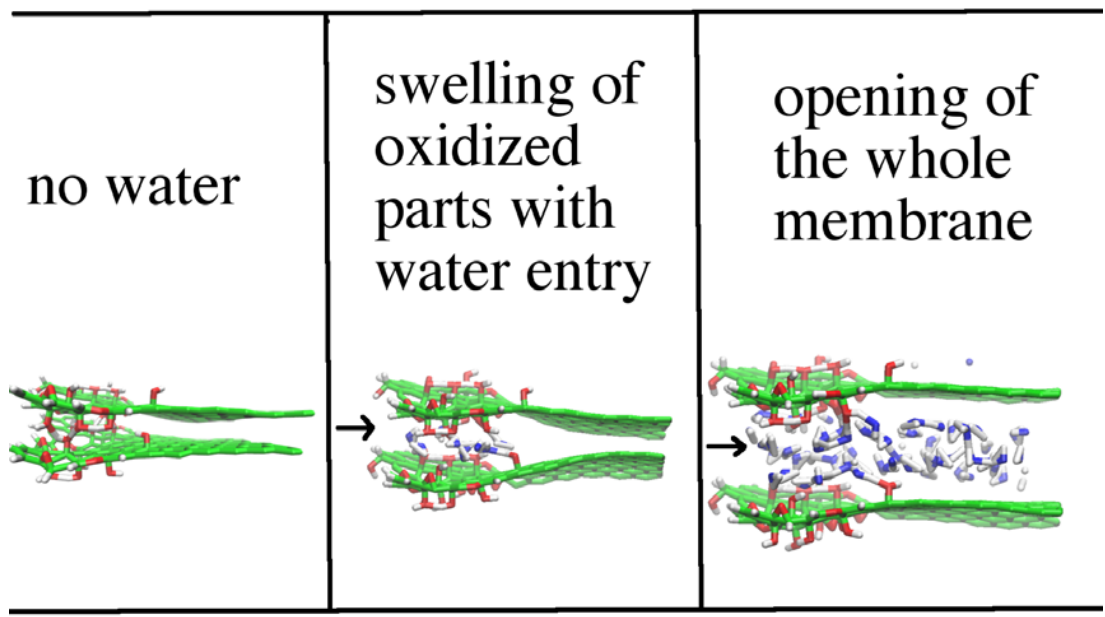

(a)

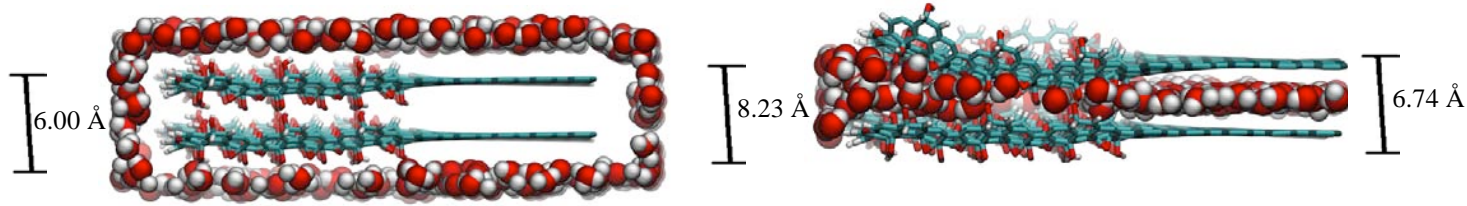

(b)

Figure 4: (a) Scheme (from left to right) showing the opening of a capillary in the presence of water. The water first gets accumulated in the oxidized parts of the membrane, and it keeps swelling it. Once it swells so much that the reduced parts of the membrane too are forced to open up. The water flow is easy through the reduced parts due to lack of hydrogen bonding of water with the membrane and due to the propensity of the reduced parts to close up to gain free energy. To distinguish between water and functionalization of graphene, water's oxygens are colored in blue and functional oxygens are colored in red. (b) Snapshots of a few representative configurations of the MD simulation of a mixed functionalized graphene sheets (60% of functionalized graphene for 33% oxidation and 40% of its reduced counterpart, area = 5 nm x 5 nm) showing the opening of a capillary in the presence of water. The water first gets accumulated in the oxidized parts of the membrane, and it keeps swelling it. Once it swells so much that the reduced parts of the membrane too are forced to open up. The water flow is easy through the reduced parts due to lack of hydrogen bonding of water with the membrane and due to the propensity of the reduced parts to close up to gain free energy.

TABLES.

**Table 1.** The translational and the rotational entropies of water trapped between two graphene sheets at different inter-sheet spacing.

| Separation in Å | $TS_{trans}$ (kcal/mol) | $TS_{rot}$ (kcal/mol) | U (kcal/mol) | U-TS (kcal/mol) | Diffusion Coefficient ($\times 10^{-5}$ cm$^2$/s) |
|---|---|---|---|---|---|
| 6.5 | 3.3 $\pm$ 0.03 | 1.14 $\pm$ 0.01 | -9.50$\pm$ 0.06 | -13.95$\pm$ 0.1 | 1.02$\pm$ 0.13 |
| 8 | 3.89 $\pm$ 0.01 | 1.14 $\pm$ 0.01 | -9.48 $\pm$0.01 | -14.51 $\pm$ 0.03 | 2.43$\pm$ 0.27 |
| 9 | 3.73 $\pm$ 0.04 | 1.05 $\pm$ 0.02 | -9.56 $\pm$ 0.07 | -14.34 $\pm$ 0.13 | 3.17$\pm$ 0.39 |
| 10 | 3.71 $\pm$ 0.03 | 1.01 $\pm$ 0.01 | -9.78 $\pm$0.004 | -14.50 $\pm$0.08 | 2.33$\pm$ 0.29 |
| 12 | 3.82 $\pm$ 0.03 | 1.01 $\pm$ 0.03 | -9.59 $\pm$0.05 | -14.41 $\pm$ 0.11 | 2.97$\pm$ 0.15 |
| 15 | 3.87 $\pm$0.03 | 0.98 $\pm$ 0.01 | -9.66 $\pm$ 0.01 | -14.51 $\pm$0.05 | 3.50$\pm$ 0.37 |
| bulk | 4.03 $\pm$0.01 | 0.95$\pm$ 0.01 | -9.57 $\pm$ 0.07 | -14.55 $\pm$ 0.09 | 6.22$\pm$ 0.90 |

**Table 2:** The translational and the rotational entropies of water trapped between two graphene oxide sheets at different inter-sheet spacing.

| Separation in Å | $TS_{trans}$ (kcal/mol) | $TS_{rot}$ (kcal/mol) | U (kcal/mol) | U-TS (kcal/mol) | Diffusion Coefficient ($\times 10^{-5}$ cm$^2$/s) |
|---|---|---|---|---|---|
| 7.0 | 2.83 ± 0.02 | 0.92 ± 0.01 | -10.50 ± 0.05 | -14.25 ± 0.08 | 0.16 ± 0.03 |
| 8.0 | 3.00 ± 0.03 | 0.86 ± 0.01 | -10.79 ± 0.08 | -14.65 ± 0.12 | 0.38 ± 0.06 |
| 9.0 | 3.04 ± 0.02 | 0.88 ± 0.01 | -10.60 ± 0.02 | -14.52 ± 0.05 | 0.47 ± 0.04 |
| 10.0 | 3.20 ± 0.01 | 0.91 ± 0.01 | -10.43 ± 0.05 | -14.54 ± 0.07 | 0.73 ± 0.03 |
| 12.0 | 3.38 ± 0.03 | 0.90 ± 0.01 | -10.29 ± 0.01 | -14.57 ± 0.05 | 1.23 ± 0.01 |
| 15.0 | 3.52 ± 0.01 | 0.92 ± 0.01 | -10.11 ± 0.01 | -14.55 ± 0.03 | 1.91 ± 0.12 |
| bulk | 4.03 ± 0.01 | 0.95 ± 0.01 | -9.57 ± 0.07 | -14.55 ± 0.09 | 6.22 ± 0.90 |